# A Novel MOSFET based Single Event Latchup Detection, Current Limiting & Self Power Cycling circuit for Spacecraft systems

Ishan Pandey, Kinshuk Gupta, Vinod Kumar, A.R. Khan and Sandhya V. Kamat

*Components Management Group*
*U R Rao Satellite Centre*
*Indian Space Research Organisation, Bengaluru-560017*

{Corresponding author's email: ishanp@ursc.gov.in}

**Abstract** - Single Event Latch-up (SEL) is one of the prime concerns for CMOS ICs used in space systems. Galactic Cosmic Rays or Solar Energetic Particles (SEP) may trigger the parasitic latch up circuit in CMOS ICs and cause increase in current beyond the safe limits thereby presenting a threat of permanent failure of the IC. Mitigation of the SEL is always a challenging task. The conventional mitigation approaches inherently introduce some response time which presents an uncertainty because during this response time the current may exceed the safe current limits. This paper presents a novel circuit based on MOSFETs which provides end-to-end complete solution of detecting SEL, limiting the current below the set threshold and executing power cycling to restore the normal functioning of the CMOS IC. The proposed circuit has been simulated in MULTISIM and the simulation results match very well with the expected behavior of (i) current limiting and (ii) the total time duration taken in power cycling to bring the SEL sensitive device back to its normal operational state. This circuit can be harnessed by spacecraft system designers to overcome the catastrophic threat of SEL posed by space radiation environment.

INTRODUCTION

The CMOS ICs contain a parasitic structure consisting of PNP and NPN transistors stacked next to each other as shown in Fig.1. Because of the PNP and NPN stacking, the parasitic structure can be considered equivalent to Silicon Controlled Rectifier (SCR). The energetic charged particles from Galactic Cosmic Rays (GCR) or Solar Energetic Particles (SEP) may trigger the SCR. If GCR or SEP particles hit near base region of the parasitic transistor (say Q2 in Fig.1), it starts conducting and hence path of current establishes for base current of transistor Q1. As base of Q2 is connected to collector terminal of Q1 and base of Q1 is connected to collector of Q2, Q1 drives Q2 and vice versa with increase in base and collector currents. They both keep each other in saturation for as long as the structure is forward biased. A low impedance path is formed between power and ground terminals which leads to continuous current increase. If current is not controlled, it may exceed the safe limit of the device causing IC damage. [1]

This phenomenon of activating this parasitic SCR structure in CMOS ICs by the strike of Galactic Cosmic Ray or Solar Energetic Particles is called Single Event Latchup (SEL).

In order to recover the normal functionality of device, it is required to turn OFF the SCR. The current through SCR should be made less than the holding current to turn it OFF. It can be done by reducing the voltage across SCR. Therefore, power cycling on detection of SEL event is required to address SEL.

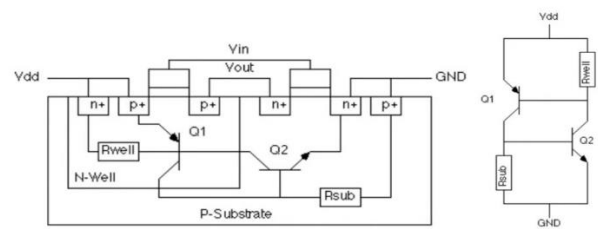

Fig.1 (a) Parasitic SCR structure in CMOS IC [2]

PREVALENT SEL MITIGATION SCHEMES AND THEIR DRAWBACKS

In space systems, the best approach to deal with SEL is to fabricate the device on insulating substrates i.e. Silicon on Insulator (SOI) technology. [3] However, SOI technology is known to have cost and yield implications. Therefore, the space system designers are constrained to employ mitigation techniques for the SEL susceptible ICs.

Short circuit protection using Fuse is a conventional technique for protecting circuits against excessive current flow. In this technique, the fuse blows off and breaks the circuit which limits the further rise of current and hence the circuit is protected. However, this technique is not useful for SEL mitigation in spacecraft systems because once blown, the fuse can't be recovered automatically and moreover it requires certain energy to be dissipated in order to get it blown, which may take some time.

The other conventional practice of current limiting is done by putting certain value of resistance in series with the device. As SEL event is triggered, current limiting resistor drops voltage across itself, causing limited current flowing through the device. The major drawback of this method is continuous power dissipation. Even for normal operating device, current has to pass through series resistor which in turn dissipates continuous power.

There is another prevalent method to mitigate SEL which is based on sensing the rise in current and if current exceeds the particular threshold, circuit breaks OFF for power

_

recycling. However, the response time of the circuit may be long and IC may experience excessive current during this response time, hence, posing a risk of damage to the device as shown in Fig.2. Therefore, for more robustness it is required to improve the circuit with the feature of limiting the current to a particular value in such a way that power consumption in normal operation must not increase, together with functionality of power recycling. The value of current limit can be kept little lower than the absolute maximum allowable current value

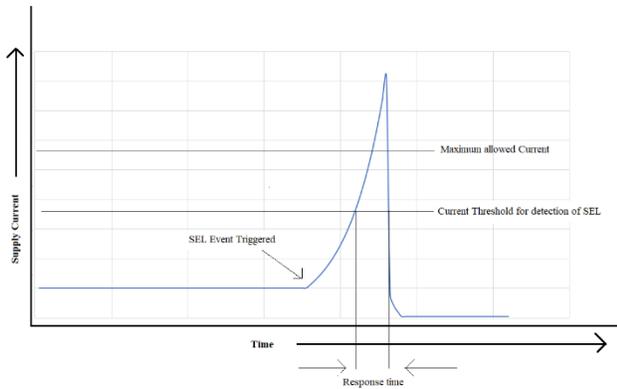

Fig.2 Transient of current passing through simulated device during SEL event without current limiter

NOVEL MITIGATION CIRCUIT TO DETECT SEL, LIMIT CURRENT AND EXECUTE SELF POWER CYCLING

The circuit presented in this paper employs the current limiting and switching characteristics of MOSFETs to build an SEL mitigation circuit which does not allow the current to increase beyond a set threshold thus eliminating the concern of response time and achieves complete power cycling to restore the normal function after the SEL event.

The circuit presented in this paper provides the complete end-to-end SEL mitigation solution for spacecraft systems by providing the following features:
1. Limit the supply current during SEL event and thus providing the assurance that the device to be protected never enters the high current state.
2. Detection of the SEL event.
3. Switch off the SEL prone device, Subsequent to latchup detection.
4. Restore its normal function after a set interval of time.

*A. MOSFET current limiting and switching characteristics*

MOSFET can operate in three regions i.e. Ohmic region, Saturation region and Cut-off region. In ohmic region, it is conducting and shows very less resistance while in saturation region, drain-source current saturates to a certain value. Value of voltage across Gate-Source ($V_{GS}$), decides the saturation current limit of the MOSFET. In this region, negligible rise in current is observed even upon increase in Drain – Source Voltage ($V_{DS}$). As the value of $V_{GS}$ increases, the value of saturation current also increases. Third region of operation is Cut-off region, MOSFET in this region does not conduct and shows very high resistance. If Gate – Source voltage is less than threshold voltage (i.e. $V_{GS} < V_T$), it operates in Cut-off region.

As we know Drain to Source current of MOSFET gets saturated if $V_{DS} \geq (V_{GS}-V_T)$. The value of Drain- Source Current ($I_{DS}$) is given by following equation (neglecting Channel Length Modulation): -

$$I_{DS} = \frac{K(V_{GS}-V_T)^2}{2} \quad \text{where } K = \frac{\mu C_{ox} W}{L} \quad (1)$$

K and $V_T$ depends on the type of MOSFET and for any given value of applied $V_{GS}$, current will not exceed $I_{DS}$. µ is the mobility of charge carrier, $C_{ox}$ is oxide capacitance per unit area of the MOSFET, W is the width of Gate and L is the length of Gate.

*B. Proposed Mitigation Circuit*

The proposed SEL mitigation circuit is shown in Fig.3.

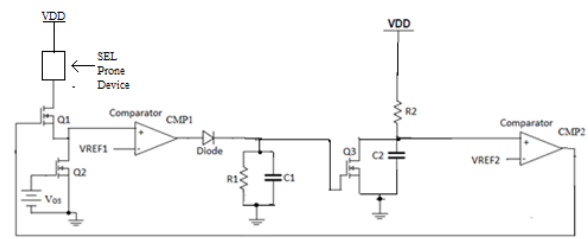

Fig.3 Proposed SEL Mitigation Circuit

The major elements of this circuit are as follows. MOSFET Q1 is used as a switch, for breaking the current path through SEL prone device. MOSFET Q2 is used as a current limiter which will not allow the current to increase beyond the set threshold value. An R1-C1 circuit consisting of parallel Resistance R1 and Capacitance C1 is used for maintaining proper turn OFF time. Comparator CMP1 is used to detect the SEL Event. MOSFET Q3 is used for providing discharge path for Capacitor C2 and Comparator CMP2 is used to control switching action of MOSFET Q1. During normal operating condition, Q1 will be turned ON, as the gate terminal is connected to output of CMP2. As it can be seen in the circuit, CMP2 compares the $V_{REF2}$ with the voltage across the capacitor C2. During normal operation of device without SEL event, MOSFET Q3 is in cut off region, so voltage across capacitor is always higher than $V_{REF2}$. Hence, output of CMP2 is always high, which drives the MOSFET Q1 in ohmic region to complete the circuit for normal operation of device.

*C. Current limiting in case SEL occurs*

During normal operation, Q2 is always turned ON with fixed value of $V_{GS}$ and voltage on Drain of Q2 is very small. Fixed value of Gate-Source voltage ($V_{GS}$) is provided on Q2 for preventing current overshoot during SEL event. The value of $V_{GS}$ can be determined from (1), where $I_{DS}$ shall be the maximum safe operating current of the device to be protected. The effect of temperature also needs to be considered because value of $V_T$ varies with temperature. Threshold voltage ($V_T$) increases as temperature decreases and vice versa. Assuming in the possible range of temperature, the threshold voltage varies between $V_{Tmin}$ to $V_{Tmax}$ and absolute maximum allowed current for the device is $I_{max}$ while max operating current is $I_{OP}$. The value of applied $V_{GS}$ should follow the following inequality.

$$V_{Tmax} + \sqrt{\frac{2*I_{OP}}{K_n}} < V_{GS} < V_{Tmin} + \sqrt{\frac{2*I_{max}}{K_n}} \quad (2)$$

The $V_{GS}$ value based on the above equation will assure that the SEL prone device never enters the high current state and hence minimize the reliability risks associated with high current operations even for small duration of time.

*D. SEL detection*

During normal operation, as Q1 and Q2 are in ohmic region, Hence, they offer minimum voltage drop across them. However, in case of SEL event, current shoots up to the saturation current limit of Q2 and it enters in the saturation region. Hence, voltage drop across drain-source of Q2 increases. As the voltage $V_{GS}$ for Q1 is kept higher than $V_{GS}$ of Q2 hence Q1 always operate in ohmic region. This increase in drain voltage is employed as a signature for detecting the SEL event. The drain voltage of Q2 is compared with a reference voltage $V_{REF1}$. $V_{REF1}$ value should be kept such that increased drain voltage of Q2 during SEL event becomes greater than $V_{REF1}$.

On occurrence of SEL event, Voltage on drain of Q2 rises above $V_{REF1}$ and hence output voltage of comparator CMP1 goes high. The CMP1 HIGH signal is fed to the follow on self power cycling circuitry. The CMP1 HIGH signal is not maintained for long because Q1 will be switched off after some duration of time. The mechanism of Q1 turning OFF is described in the subsequent sections. Thus, CMP1 provides a temporary HIGH signal to the follow on circuitry to initiate the self power cycling mode.

*E. Self Power Cycling*
*(i) Turning the SEL prone device OFF*

The high-level output voltage of CMP1 instantly charges the capacitor C1. Gate of MOSFET Q3 is connected with capacitor C1. Q3 is used as a switch, which provides the discharging path for capacitor C2. (purpose of C2 is described in subsequent section). As voltage across C1 rises above the threshold voltage of MOSFET Q3, it turns ON.

Capacitor C2 is connected across drain – source of Q3. C2 and R2 are connected in series with supply voltage to provide fixed amount delay before resuming power supply to SEL prone device. During long operation of device without SEL event, C2 gets charged completely to $V_{DD}$. Due to absence of discharging path, C2 withholds voltage $V_{DD}$ constantly. But as described above, in case SEL event occurs, Q3 turns ON and hence capacitor C2 gets the discharging path across drain-source of Q3. During discharging event of C2, the voltage across C2 keeps on decreasing. The voltage across C2 is fed to comparator CMP2. It is used for disconnecting and resuming the power supply to the SEL prone device. It compares the voltage across C2 with a reference voltage $V_{REF2}$. Value of $V_{REF2}$ can be any value greater than zero but less than $V_{DD}$. As the voltage across C2 starts decreasing, there comes a point where it becomes less than $V_{REF2}$ and the output of CMP2 goes low, causing MOS Q1 to turn OFF, which results in break of circuit and SEL prone device turns OFF. The device will remain OFF until Q1 is OFF which implies until Voltage across C2 is more than $V_{REF2}$.

*(ii) Maintaining appropriate turn off time*

Main function of parallel R1 & C1 with diode is to ensure proper turn off time is given to the circuit. As output of CMP1 goes high during first SEL event, Q3 turns ON with small drain source resistance $R_{DS(ON)}$. This drain source resistance does not allow capacitor C2 to get complete discharge instantaneously. The time obtained for C2 to discharge across Q3 would be dependent on propagation delay of CMP2 & CMP1 and turn ON time of MOSFET Q1. As voltage of C2 goes just below $V_{REF2}$, CMP2 generates zero in its output which turns off Q1 and hence CMP1 generates low on its output disrupting the discharging path of C2. After this point C2 starts to get charging from the voltage where discharging stopped. Therefore, to obtain proper turn OFF time parallel R1- C1 is used. Diode is used to prevent backflow of current from capacitor C1.

After powering off the circuit, time taken to restart the power is given by: -

*Total time to restart= Time duration for which Q3 remain ON (T1) + Time taken to charge C2 to voltage more than $V_{REF2}$ (T2)*

Assuming maximum voltage across capacitor C1 (i.e. maximum output voltage of CMP1 considering the drop across diode D1) is $V_o$,

$$T_1 = R1 * C1 * \ln\left(\frac{V_o}{V_T}\right)$$

And,

$$T_2 = R2 * C2 * \ln\left(\frac{V_{DD}}{V_{DD} - V_{REF2}}\right)$$

*Total time for which device remains turned off*
$$= R1 * C1 * \ln\left(\frac{V_o}{V_T}\right) + R2 * C2 * \ln\left(\frac{V_{DD}}{V_{DD} - V_{REF2}}\right) \quad (3)$$

*(iii) Restarting the device*

After disrupting the power supply to SEL prone device, circuit waits for certain amount of time. This wait time as described above is the combination of time in which Q3 MOSFET is ON and the time taken by C2 to get charged to more than $V_{REF2}$ via R2 resistor. As voltage across C2 gets more than $V_{REF2}$, CMP2 would generate high in its output which turns ON the MOSFET Q1. Hence, resuming the power supply to the device.

SIMULATION MODEL

The circuit presented above is simulated on MULTISIM software using PSPICE model of components present in its master database. The components selected for simulation are presented in Table-1 and the simulation diagram is shown in Fig.4.

Table-1

| Component | Selections for Simulation |
|---|---|
| Q1, Q2, Q3 | 2N7000 |
| VREF1, VREF2 | 5 V |
| CMP1, CMP2 | AD8561AR |
| Diode | 1N4454 |
| R1, R2 | 10kΩ |
| C1 | 1μF |
| C2 | 10μF |

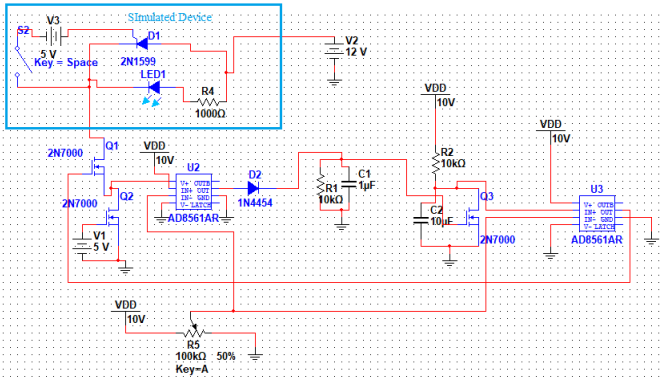

Fig.4 Simulation circuit in Multisim

### A. Simulation of SEL Prone device on Multisim

Device to be protected is simulated as LED in series with resistor and an SCR parallel to them depicting the parasitic circuit inside device. SEL event is simulated by firing the SCR through gate bias with help of key. To demonstrate that the proposed circuit can give protection to a wide variety of SEL prone devices, four SEL threshold currents are selected for this device viz. 450 mA, 150 mA, 50 mA and 12 mA and separate simulations have been carried out for each case. Current probe is connected to read the value of current flowing through the simulated device with 1mV/mA setting.

### B. Biasing of devices

Simulated SEL prone device is biased with 12V, which is in series with MOSFET Q1 and MOSFET Q2. Corresponding to the four cases of SEL current thresholds as described in previous section, the $V_{GS}$ values of MOSFET Q2 are 5V (for 450 mA), 4V (for 150 mA), 3V (for 50 mA) and 2.5V (for 12 mA). Comparator IC, AD8561 is connected with 10 V supply and reference voltages ($V_{REF1}$ & $V_{REF2}$) are derived using POT to 5V.

### RESULTS

### A. Current limiting

On simulating the SEL event on MULTISIM, for the four SEL threshold current cases as described above, it is observed that the current through the device gets limited to the SEL threshold current values thus protecting the device from high current state. It is also observed that the response time of circuit to switch OFF the device is not more than 50 μsec. The results are shown in Fig.5.

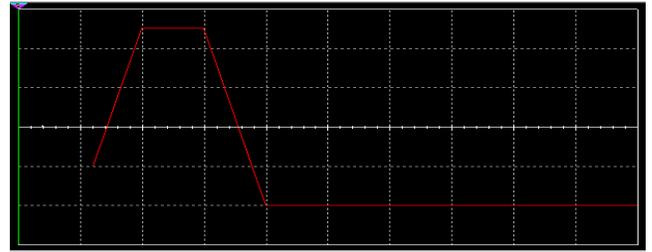
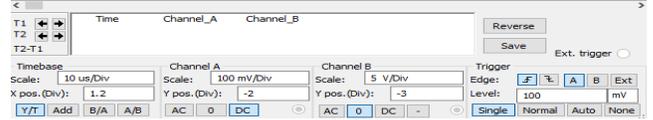

$V_{GS}$=5 V, Max Current 450 mA

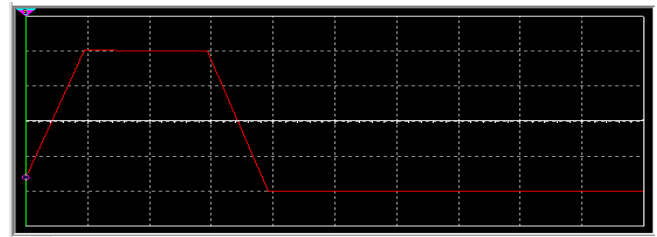
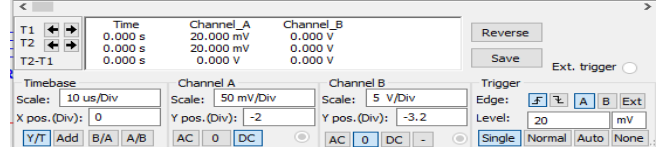

$V_{GS}$=4 V, Max Current 150 mA

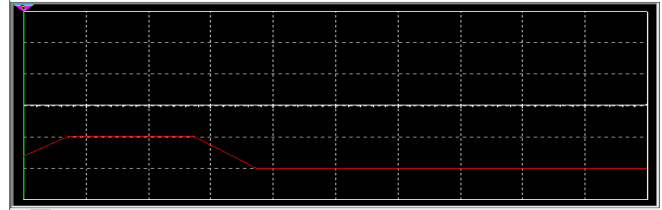
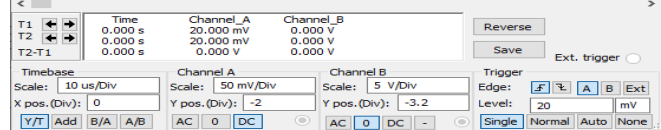

$V_{GS}$=3 V, Max Current 50 mA

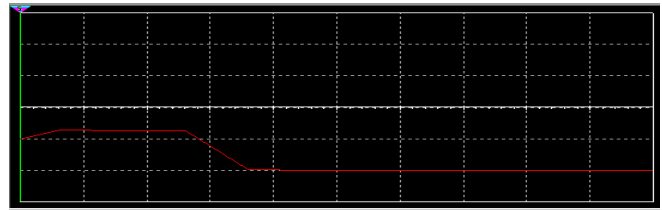
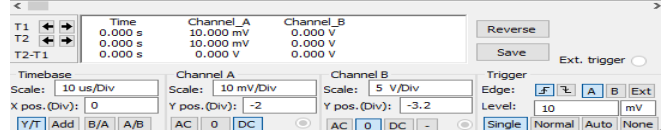

$V_{GS}$=2.5 V, Max Current 12 mA

Fig.5 Transient of current passing through simulated device during SEL event.

### B. SEL Detection and Self Power Cycling

C. After a long time of normal operation without SEL event C2 gets charged to 10V and voltage across C1 capacitor would be zero. After SEL event detected, C1 gets charged which creates the discharging path for C2 via MOS Q3. Hence, C2 discharges instantly and then starts charging after the voltage across C1 decreases below threshold voltage of Q3 MOS as shown in Fig.6.

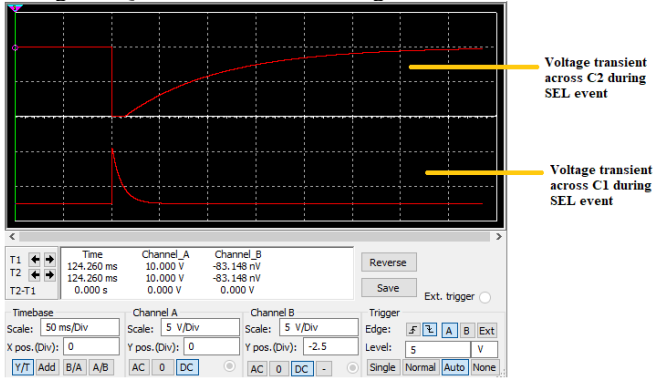

Fig.6 Transient behaviour of voltage across capacitors C1 & C2 during SEL

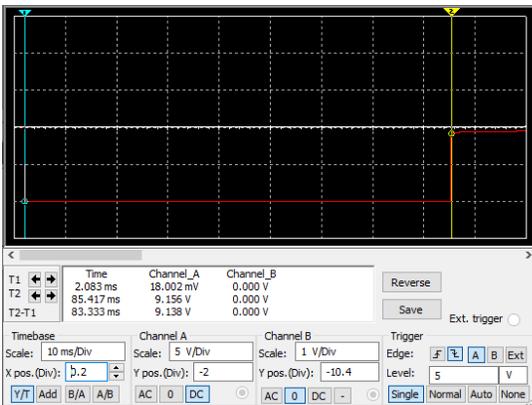

Fig.7 Transient of Gate voltage of Q1 during power recycling on SEL event

Comparison is done for the total time during which the device remains turned off, obtained from simulation results to that of theoretical estimation using (3), From the simulation result shown in Fig.7, it is observed that the turn off time is 83.33 millisecond. Maximum output voltage of comparator CMP 1 ($V_o$) considering the drop across Diode D1 is observed to be 8.8 V and threshold voltage ($V_T$) of MOSFET Q3 is 2.1 V. Putting these values along with the other values of R & C of the circuit, in (3), *Total time for which device remain turned off* comes out to be 83.645 millisecond which matches very well with the value obtained from simulation.

### DISCUSSION

The results show that the proposed circuit achieves the goal of limiting the current below set threshold and also initiate automated power cycling operation. This circuit can be used for any SEL sensitive device which is required to be protected. Based on the safe current limit, either $V_{GS}$ can be changed or if $V_{GS}$ is to be kept fixed because of other design constraints, then MOSFET Q2 can be chosen such that the threshold voltage of MOSFET could help in maintaining saturation current limit as per equation (1).

Some devices may require to be kept in OFF state for longer durations so as to ensure that the Latchup event has died down. This circuit provides the feature to tune the time for which the device is to be kept in OFF mode after detecting an SEL event by appropriate selection of R1, R2, C1 and C2. Thus, this design offers great flexibility to space system designers and it can be used for any SEL sensitive device protection.

### CONCLUSION

This paper has presented a novel circuit design to address the problem of Single Event Latchup (SEL) in CMOS ICs used in spacecraft systems. The proposed circuit is a comprehensive solution which not only keeps the SEL sensitive device always below the set current threshold thus protecting it but also achieves automatic fast recovery of the device to its normal operating state. Compared to other prevalent techniques to deal with SEL, this design provides absolute certainty that the SEL sensitive device will never enter the high current state thus providing reliability assurance for the space systems. The simulation results show that even if SEL event occurs, the loss in functionality is few tens of milliseconds. Hence this circuit provides a very high degree of assurance for spacecraft service availability even in extremely harsh space radiation environments.